\documentclass{emulateapj}

\usepackage{graphicx}
\graphicspath{{plots/}}

\usepackage{verbatim}
\usepackage{amsmath}
\usepackage{multirow}

\usepackage{epstopdf}
\usepackage{color,soul}

\usepackage[pdftitle={AGN Heating and the Low-redshift Ly$\alpha$ Forest},pdfauthor={Alex Gurvich}]{hyperref}

\newcommand{\km}{\mathrm{km}}

\newcommand{\s}{\mathrm{s}}

\newcommand{\NHunit}{cm$^{-2}$}

\newcommand{\Mpch}{\, h^{-1} \mathrm{Mpc}}

\newcommand{\NHI}{\mathrm{N}_\mathrm{HI}}
\newcommand{\gadget}{GADGET}
\newcommand{\arepo}{AREPO}
\newcommand{\kms}{\ensuremath{\km\s^{-1}}}
\newcommand{\Lya}{Lyman-$\alpha$}

\shorttitle{AGN Heating and the Low-redshift Ly$\alpha$ Forest}

\shortauthors{Gurvich et al.}

\begin{document}

\title{The Effect of AGN Heating on the Low-redshift Ly$\alpha$ Forest}

\author{Alex Gurvich\altaffilmark{1}, 
Blakesley Burkhart\altaffilmark{2} \& Simeon Bird\altaffilmark{3}}
\altaffiltext{1}{Department of Physics and Astronomy and CIERA, Northwestern University, 2145 Sheridan Road, Evanston, IL 60208, USA \\
 \href{mailto:agurvich@u.northwestern.edu}{\nolinkurl{agurvich@u.northwestern.edu}}}
\altaffiltext{2}{Harvard-Smithsonian Center for Astrophysics, 60 Garden Street, Cambridge, MA 0213, USA }
\altaffiltext{3}{Department of Physics and Astronomy, Johns Hopkins University, 3400 N. Charles Street, Baltimore, MD 21218, USA}

\begin{abstract}
We investigate the effects of AGN heating and the ultraviolet background on the low-redshift Lyman-$\alpha$ forest column density distribution (CDD) using the Illustris simulation.
We show that Illustris reproduces observations at $z =0.1$ in the column density range $10^{12.5} - 10^{13.5}$cm$^{-2}$, relevant for the ``photon underproduction crisis.'' 
We attribute this to the inclusion of AGN feedback, which changes the gas 
distribution so as to mimic the effect of extra photons, as well as the use of the Faucher-Gigu\`ere ultraviolet background, 
which is more ionizing at $z=0.1$ than the Haardt \& Madau background previously considered.
We show that the difference between simulations run with smoothed particle hydrodynamics and simulations using a moving mesh is 
small in this column density range but can be more significant at larger column densities. 
We further consider the effect of supernova feedback, Voigt profile fitting and finite resolution, all of which we show to have little influence on the CDD.
Finally, we identify a discrepancy between our simulations and observations at column densities $10^{14} - 10^{16}$cm$^{-2}$, where 
Illustris produces too few absorbers, which suggests the AGN feedback model should be further refined. 
Since the ``photon underproduction crisis'' primarily affects lower column density systems, we conclude that AGN feedback and standard ionizing background models can resolve the crisis.

\end{abstract}


\section{Introduction}
\label{intro} 

Fifty years after \cite{Gunn:1965} and \cite{Bahcall:1965} used quasar absorption systems to
infer the neutral hydrogen density, the \Lya~forest remains a key diagnostic of
galaxy physics and cosmology. \Lya~forest observations at $z \geq 2$, where the \Lya~line 
can be observed from the ground, have been used to constrain the small-scale cosmic 
structure \citep{Croft:1998, McDonald:2004pk}, the temperature of dark 
matter \citep{Viel:2009}, the gas temperature \citep{Becker:2011, Boera:2014} and the evolution of the metagalactic ionizing 
background \citep{HaardtMadau:1996, FG_2008, FG_Lidz:2008, Faucher:2009, HaardtMadau:2012}.
Meanwhile, observations of the abundance of stronger neutral 
hydrogen absorbers have been used as a test of cosmological structure models
\citep[e.g.~][]{Cen:1994, Zhang:1995, Hernquist:1996, Miralda:1996, Rauch:1997, Pontzen:2008,FG_DK:2011, Fumagalli:2011, Fumagalli:2014, Bird:2014, FG_LLS:2015,FG_LLS:2016}.
Studying the diffuse absorbing gas that produces the \Lya~forest requires cosmological hydrodynamic simulations, which include both fully nonlinear 
gravitational collapse and the interaction between gravitational and photoionization heating and cooling processes.

Recent observations with the Hubble Space Telescope  \citep[\textit{HST};][]{Lehner:2007, Danforth:2016} have enabled 
studies of the low-redshift \Lya~forest. While at $z=2$ the \Lya~forest arises from an approximately 
mean density gas, by $z=0$ it is a probe of gas closer to $10$ times the cosmic mean \citep{Dave:1999}.
Thus, the low-redshift \Lya~forest traces the diffuse gas in filaments and the outskirts of galaxies and clusters.

Importantly, photoionization equilibrium between the neutral hydrogen and the ionizing background provides a unique 
census of the number of ionizing photons at $z \approx 0$.
In particular, Kollmeier et al. (\citeyear[][hereafter K14]{Kollmeier:2014}) reported a discrepancy of a factor of $3.3$ between 
the observed column density distribution (CDD) data of \cite[][D16]{Danforth:2016}, taken with the
Cosmic Origins Spectrograph (COS) on HST, and the CDD derived using synthetic \Lya~spectra 
from a simulation run with \gadget. They suggest that 
resolving this discrepancy requires increasing the number of ionizing photons at $z \approx 0$ by 
a factor of five over their fiducial \cite[][HM12]{HaardtMadau:2012} ultraviolet background (UVB) model. They consider this increase
unlikely to be consistent with current observational uncertainties in the escape fraction of 
ionizing photons from galaxies. Thus, they suggest that resolving 
this ``photon underproduction crisis'' may require previously 
unobserved heating sources, such as blazars \citep{Chang:2012, Puchwein:2012} or annihilating dark matter particles.
Later work \citep{Khaire:2015} suggested that enough photons could be produced using updated models 
of quasar emissivity. \cite{Shull:2015} suggested that the discrepancy could be resolved using 
only a factor of two increase in the ionization rate $\Gamma_{HI}$, a difference that they considered to be plausible.

Here, we re-examine this discrepancy in the context of the Illustris, a cosmological hydrodynamic 
simulation using the moving-mesh code \arepo, which includes a comprehensive model for the evolution of gas and galaxies to $z=0$, and has 
been shown to match several low-redshift properties of galaxies \cite[e.g.~][]{Genel:2014, Vogelsberger:2014, Sijacki:2015}.
We compare the CDD of neutral hydrogen in the \Lya~forest from Illustris to the observations of D16. We investigate the 
effects of supernova and AGN feedback prescriptions, the hydrodynamic solver, the UVB model, and the method 
for estimating column densities. We find that the previously neglected effect of AGN feedback, together with a $1.7\times$ increase in the ionization rate over HM12 due a different UVB model from \cite{Faucher:2009}, is 
sufficient to resolve the discrepancy at low column densities.

The paper is organized as follows. In Section~\ref{sec:sims}, we introduce the simulations, in Section~\ref{sec:res},
we present the results of the comparison of the CDD from the simulations to the observations of D16. 
In Section~\ref{sec:dis}, we discuss our results, followed by our conclusions in Section~\ref{sec:con}.

\section{Simulations}
\label{sec:sims}

We use the Illustris simulation \citep{Genel:2014, Vogelsberger:2014, Nelson:2015} to analyze the low-redshift 
\Lya~forest CDD. Illustris is a cosmological hydrodynamic simulation in a box of length $75 \Mpch$. Gravitational interactions 
from dark matter and baryons are evolved using the TreePM algorithm \citep{Springel:2005}.
Radiative cooling is implemented using a rate network following \cite{Katz:1996}, including line cooling,
free-free emission, and inverse Compton cooling.  Illustris assumes ionization equilibrium and accounts for shielding 
from the radiation background at high hydrogen column densities, following \cite{Rahmati:2013a}. Shielding 
is followed during the course of the hydrodynamic calculation and is thus included in the dynamics of the simulation. Metals and metal-line cooling are 
included as described in \cite{Vogelsberger:2012}. star-formation is implemented using the subgrid model of \cite{Springel:2003}. 
The star-forming gas is assumed to have a temperature of 
$\sim 10^4$K and is thus fully neutral for the purposes of HI absorption.

For hydrodynamics, Illustris uses the moving-mesh code \arepo~ \citep{Springel:2010}.
Each grid cell on the moving mesh is sized to contain a roughly fixed amount of mass and to move approximately following
the local bulk motion of the fluid. Small-scale mixing is included by allowing gas and metals to advect between grid cells.
To assess the impact of the hydrodynamical solver on the CDD in Illustris, we examined two simulations 
from \cite{Vogelsberger:2012}. While these simulations were designed to have identical initial conditions and gravitational evolution, 
they used different codes with different hydrodynamical solvers. One used \arepo, while the other used the smoothed particle 
hydrodynamics (SPH) code \gadget -3 \citep{Springel:2005}. Neither of these simulations include feedback, allowing
a clean comparison between numerical hydrodynamical methods without being affected by implementation differences in the star-formation/AGN models.

The ultraviolet background (UVB) in Illustris follows the estimates of Faucher-Gigu\`ere  (\citeyear[henceforth FG09]{Faucher:2009}).
K14 used the UVB model of HM12. Both UVB models are calibrated primarily at $z=2-4$, 
but the FG09 UVB has a shallower slope at lower redshift, so that by $z \sim 0$ it produces about $1.7$ times more ionizing photons, 
differing from HM12 by $\Delta \Gamma_{HI}= 1.6 \times 10^{-14}$ s$^{-1}$.
To confirm the magnitude of this effect on the CDD, we have performed a simulation using the Illustris fiducial setup, but with an HM12 UVB.

As described in detail in~\cite{Vogelsberger:2013}, Illustris includes phenomenological models for stellar and AGN feedback
that aims to capture the unresolved influence of these energetic events on their environment. The parameters of these feedback models 
have been adjusted to approximately reproduce the galactic stellar mass function and 
star-formation rates at $z=0$ by suppressing 
star-formation relative to pure gravitational collapse. In general, the supernova feedback model dominates in low-mass objects, while
the AGN feedback is effective in high-mass systems. 

The supernova feedback model suppresses star-formation via kinetic feedback from star-forming cells to nearby gas cells. 
The total energy of the supernova wind is held constant and is given by
\begin{equation}
 \mathrm{egy}_\mathrm{w} = \frac{1}{2} \eta_\mathrm{w} v_\mathrm{w}^2\,,
 \label{eq:energy}
\end{equation}
where $\eta_\mathrm{w}$ is the wind mass loading and $v_\mathrm{w}$ is the
wind velocity. $v_\mathrm{w}$ scales with the local dark matter velocity dispersion,
which correlates with the maximum dark matter circular velocity of the host halo \citep{Oppenheimer:2008}.
The Illustris wind model thus yields large mass loadings in small halos (as $\mathrm{egy}_\mathrm{w}$ is constant), which
allows it to roughly match the galaxy stellar mass function at $z=0$ \citep{Okamoto:2010, Puchwein:2013}. The parameters of
the wind model are described in detail in \cite{Vogelsberger:2013}.

As implemented in Illustris, AGN feedback suppresses star-formation in the most massive halos by periodically releasing thermal energy
from the black hole into the gas cells surrounding it \citep{DiMatteo:2005, Springel:2005f,Sijacki:2007, Vogelsberger:2013}.
The AGN feedback model has two modes: quasar-mode feedback and radio-mode feedback. Quasar-mode feedback is in operation when
the central black hole of the halo has high accretion rates. Here, a small fraction of the rest-mass energy of the accreted
material couples directly to the dense gas surrounding the black hole. Radio-mode feedback, which operates at low 
accretion rates, is implemented by the formation of thermal bubbles in the gas around the black hole. This feedback mode 
significantly decreases the star-formation rate in massive halos \citep{Vogelsberger:2013} and substantially reduces the 
gas density up to $\sim 1$~Mpc from the host halo \citep{vanDaalen:2011, Sijacki:2015}. Illustris also includes the local effect of radiation from the 
central quasar, although this is only important for accretion rates close to the Eddington limit. 
\cite{Weinberger:2016} suggested an updated model for accretion and feedback effects of AGN to 
ameliorate several known discrepancies between observed and simulated galaxy properties in the Illustris simulation, such as the low gas fraction in groups of galaxies
and clusters.  Here, we investigate the \Lya~forest using the fiducial Illustris feedback model \citep{Vogelsberger:2013} and will 
explore the effects of the updated prescription of \cite{Weinberger:2016} in a future work.

Illustris initially contains $1820^3$ dark matter particles and 
$1820^3$ gas elements. To check the convergence of our results with respect to the box size and resolution, and to investigate the 
impact of different feedback models and hydrodynamical solvers, we used several smaller simulations, all with a box size of $25 \Mpch$.
The simulations varying the hydrodynamical solver and UVB amplitude have $2 \times 512^3$ particles, giving a resolution approximately equal to Illustris, while the simulations 
varying the feedback model have $2\times 256^3$ particles. We summarize the basic setup of all simulations used in this study and 
provide their parameters in Table \ref{tab:models}.

\begin{table*}
\begin{center}
\caption{Description of Simulations
\label{tab:models}}
\begin{tabular}{ccccccccc}
\hline\hline
Name    	& AGN Feedback? & $\epsilon_\mathrm{m}$ \footnote{$\epsilon_\mathrm{m}$ is the AGN feedback radio-mode energy fraction. See Table 1 of 
\cite{Vogelsberger:2013}.} & SNe Feedback? & UVB Model used & Box Size & Particles & Reference &Figures  \\	
\tableline
Illustris 	& Yes 		& 0.35 & Yes 		& FG09 		   & $75 \Mpch$ & $1820^3$ & V14\footnote{\cite{Vogelsberger:2014}} & 
\ref{fig:illustris},\ref{fig:resolution},\ref{fig:voigt}\\
Illustris-small & Yes 		& 0.35  & Yes 		& FG09 		   & $25 \Mpch$ & $512^3$ & V13\footnote{\cite{Vogelsberger:2013}} &  
\ref{fig:uvb},\ref{fig:resolution}\\
Illustris-lowres & Yes 		& 0.35 & Yes 		& FG09 		   & $25 \Mpch$ & $256^3$ & V13 & \ref{fig:agn},\ref{fig:feedback},\ref{fig:resolution}\\
Stellar		& No		& ... & Yes		& FG09		   & $25 \Mpch$ & $256^3$ & V13  &\ref{fig:feedback}\\
No Feedback	& No		& ... & No		& FG09		   & $25 \Mpch$ & $256^3$ & V13 &\ref{fig:feedback},\ref{fig:agn}\\
Stronger Radio 	& Yes 		& 0.7 & Yes 		& FG09 		   & $25 \Mpch$ & $256^3$ & V13 & \ref{fig:agn}\\
Weaker Radio 	& Yes 		& 0.175 & Yes 		& FG09 		   & $25 \Mpch$ & $256^3$ & V13 &\ref{fig:agn}\\
HM12$_{\rm{AREPO}}$ 		& Yes 		& 0.35 & Yes 		& HM12		   & $25 \Mpch$ & $512^3$ & ... &\ref{fig:uvb} \\
AREPO$_{\rm{test}}$ 		& No 		& ... & No 	&  FG09		& $25 \Mpch$ & $512^3$	&  V12\footnote{\cite{Vogelsberger:2012}} &\ref{fig:arepovgadget}\\
Gadget$_{\rm{test}}$ 		& No		& ... & No 	& FG09		& $25 \Mpch$ & $512^3$	&  V12 &\ref{fig:arepovgadget} 
\\
\hline
\end{tabular}
\end{center} 
\end{table*}

\subsection{Analysis}

We define the CDD function, $f(\NHI)$, by
\begin{equation}
f(\NHI)=\frac{F(N)}{\Delta N} \Delta z\,.
\end{equation}
$F(N)$ is the number of absorbers per sightline with column density in the interval
[$\NHI$;$\NHI+\mathrm{d}\NHI$], over the redshift interval $\Delta z$ contained in our simulated spectra. 
We estimate the CDD from the column density along $25,000$ simulated skewers randomly positioned in each simulation box.
Column densities are computed by interpolating the neutral hydrogen in each gas element to the sightline using an SPH kernel.
A single absorber is $50 \kms$ across, although we have checked that our results are not sensitive to this value.
Thus, our column densities correspond to the integrated physical density field in the simulation and 
provide a quick and scalable way to examine the physical state of the absorber. The procedure is described in detail 
in \cite{Bird:2014a} and the implementation is available at \url{https://github.com/sbird/fake_spectra}.

We also investigated estimating column densities by Voigt fitting to the optical depth along our simulated spectra.
We describe our procedure in Appendix \ref{sec:voigt} and show that the method used makes a negligible difference to our results.
Following K14 we perform Voigt fitting using AUTOVP~\citep{Oppenheimer:2006}.

\section{Results}
\label{sec:res}

\begin{figure}
\centering
\includegraphics[width=0.45\textwidth]{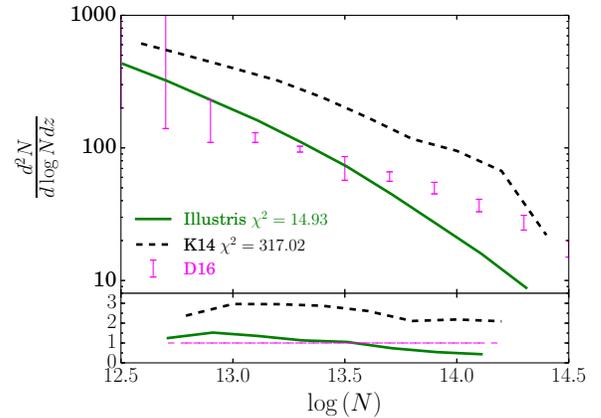}
\caption[]{
CDD from Illustris $75 \Mpch$ at $z=0.1$ (green solid), compared to the simulations of K14 \citep[black dashed]{Kollmeier:2014}
and the observations of D16 \citep[magenta points]{Danforth:2016}. The lower panel shows the ratio of each simulation with the D16 results. The column density 
range is chosen to match that shown by K14.  Illustris reproduces
the amplitude of the observed CDD well, without the offset seen by K14. However, Illustris does not reproduce the shape of the CDD at
higher column densities, $\NHI > 10^{13.5}$cm$^{-2}$, producing too few absorbers.
}
\label{fig:illustris} 
\end{figure}

Figure \ref{fig:illustris} shows our main result, the $z=0.1$ HI CDD from the Illustris simulation compared to the 
numerical results of K14 and observations of D16.  Illustris reproduces
the amplitude of the observed CDD at low column densities without the offset seen by K14. 
However, Illustris does not reproduce the shape of the CDD at
higher column densities, $\NHI > 10^{13.5}$cm$^{-2}$, producing too few absorbers.
In the rest of this section, we discuss why Illustris reproduces the observed CDD amplitude at low column densities, while leaving the discrepancy in the shape at higher column densities for Section \ref{sec:dis}.

The observational survey of D16 includes absorbers for $z=0.1-0.47$. While we report results at $z=0.1$, near the median redshift of the survey, 
we have checked that the CDD is very similar if we average simulation outputs over the observed range of redshifts.
For easy comparison with K14, we show the column density range used in that work, but in future plots we also show larger column densities, as the
observational survey of D16 includes these absorbers. 

\subsection{Effect of the Hydrodynamic Solver}

\begin{figure}
\centering
\includegraphics[width=0.46\textwidth]{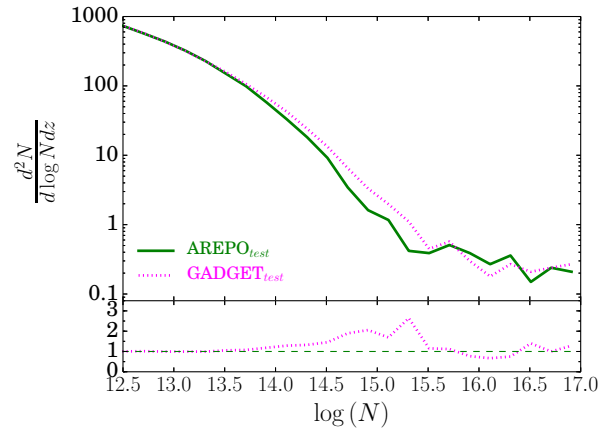}
\caption[ ]{
Effect of the hydrodynamic solver on the column density distribution at $z=0$ 
using smoothed particle hydrodynamics (\gadget, magenta dashed) and an otherwise identical simulation performed using a moving-mesh (\arepo, green solid).
The lower panel shows the \gadget~ simulation divided by the \arepo~ simulation. There is not a significant difference between the two codes at the low column density range studied in K14
}
\label{fig:arepovgadget}
\end{figure}

One possible reason for the discrepancy between our results and the simulations of K14
is the hydrodynamic solver used. Illustris uses the code \arepo~with a moving-mesh solver for hydrodynamics, 
while the simulations analyzed in K14 use \gadget~ with smoothed particle hydrodynamics.
Figure~\ref{fig:arepovgadget} shows results from the $z=0$ output of two simulations from \cite{Vogelsberger:2012} designed specifically
to assess the effect of the hydrodynamic solver. These simulations have identical initial conditions 
and gravity solvers, but different hydrodynamic methods.

For column densities $10^{14} < \NHI < 10^{15.5}$\NHunit,
\arepo~ produces about half as many absorbers as \gadget. As shown in \cite{Bird:2013},
\gadget~ overpredicts the number of $10^{17}$~\NHunit~absorbers at $z=3$ by a factor of two, and here we see
the low-redshift analogue of that discrepancy. Note that the column density $10^{15}$~\NHunit at $z \sim 0$ probes a similar 
physical density to a column density of $10^{17}$~\NHunit~at $z \sim 3$ \citep{Dave:1999}.
However, the difference between 
D16 and K14 is confined to column densities $\NHI < 10^{14}$~\NHunit, and thus the 
hydrodynamic solver cannot account for the differences in that range. 

\subsection{The UVB}

\begin{figure}
\centering
\includegraphics[width=0.46\textwidth]{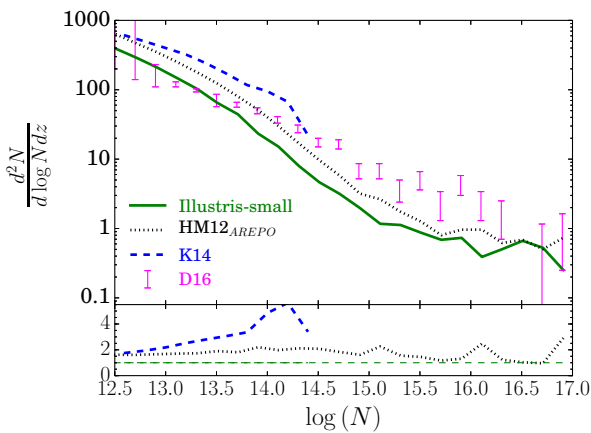}
\caption[ ]{Impact of changing the ultraviolet background (UVB) on the CDD at $z=0.1$. 
The Illustris-small simulation, which uses the FG09 UVB, is indicated in green (solid), while
an identical simulation using the HM12 UVB is shown in black (dotted).  The lower panel shows the ratio of each curve to the Illustris-small results.
The different UVBs used account for about half of the difference between the CDD in Illustris and the CDD of K14 (as shown with the HM12$_{\rm{AREPO}}$ simulation).
    }
\label{fig:uvb}
\end{figure}

Figure \ref{fig:uvb}~shows the effect of changing the UVB on two \arepo~simulations using the Illustris feedback model. 
We compare the FG09 UVB, the default in Illustris, to the HM12 UVB used by K14.
Both UVB models are calibrated primarily at $z=2-4$, but the FG09 UVB decreases less strongly at lower redshifts, so 
that by $z \sim 0$ it produces $1.7$ times more ionizing photons.
As discussed by K14, at $z \sim 0.1$ these column densities probe gas that is both highly ionized and in photoionization equilibrium.
Thus, $\NHI \propto 1/\Gamma_{HI}$, so that the effect of changing the UVB from HM12 to FG09 is to decrease column densities by $1.7$.
In practice, the effect is slightly larger than expected; the FG09 simulation is a good match to the HM12 simulation when column 
densities are divided by a factor of $1.9$, reflecting second-order terms in the photoionization equilibrium.
Thus, the different UVBs used account for about half of the difference between the CDD in Illustris/D16 and the CDD of K14.

\subsection{The Effects of Feedback}

\begin{figure}
\centering
\includegraphics[width=0.46\textwidth]{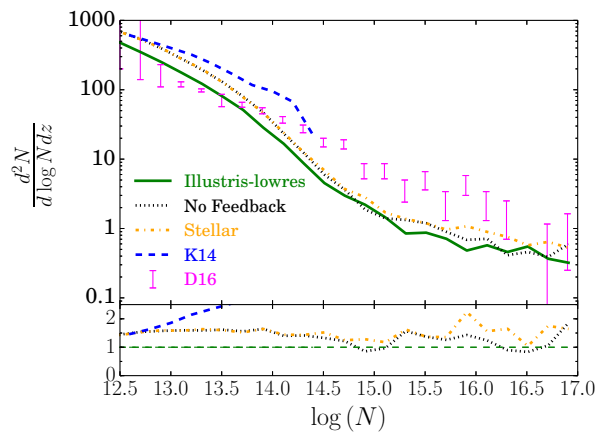}
\caption[ ]{Effect of feedback models on the $z=0.1$ CDD: the Illustris-lowres $25$ Mpc simulation at $z=0.1$, 
which includes both stellar and AGN feedback, is indicated in green (solid). A simulation with neither AGN nor supernova 
feedback is indicated with a black (dotted) line, while a similar model without AGN feedback but with supernova feedback 
is shown in blue (dot-dashed) and labeled ``Stellar.'' The lower panel shows the ratio of each CDD to the Illustris-lowres simulation. 
The supernova feedback has negligible effect on the CDD for column densities $\NHI < 10^{16}$\NHunit. }
\label{fig:feedback} \end{figure}


\begin{figure}
\centering
\includegraphics[width=0.46\textwidth]{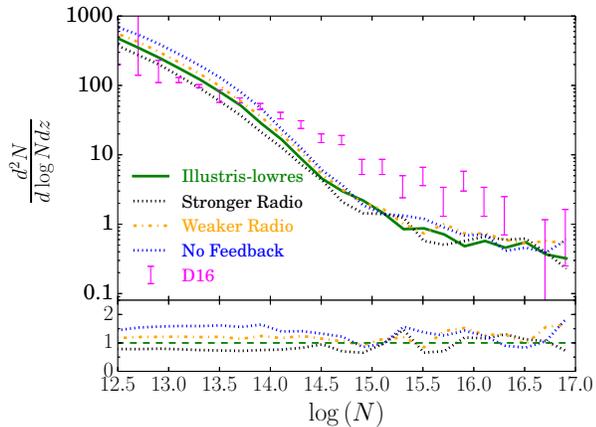}
\caption[ ]{
Effect of changing the strength of the AGN feedback from Illustris-lowres simulation (green solid line) with
stronger (black dotted line) or weaker (orange dot-dashed line) radio-mode prescriptions along with the no feedback case for reference.
The lower panel shows the ratio of each simulation column density distribution to the Illustris-lowres simulation.
}
\label{fig:agn}
\end{figure}

Figure~\ref{fig:feedback} shows the effect of the supernova feedback models used. We show a simulation with no feedback, one with 
supernova feedback only, and finally the Illustris-lowres simulation, which includes both stellar and AGN feedback.
The supernova feedback has a negligible effect on the CDD for column densities $\NHI < 10^{16}$~\NHunit. 
This is not surprising. Supernova feedback changes the distribution of the gas only in regions relatively close to 
to star-forming halos, while the CDD at this column density probes gas up to a few megaparsecs from galaxies \citep{Dave:1999, Shull:2015}. 
The simulations of K14 include supernova feedback but not AGN feedback. The ``vzw'' feedback model \citep{Oppenheimer:2008} used in K14 
is similar to ours but with parameters that make it somewhat more efficient at expelling gas from large halos.
Their results are thus most closely comparable to 
our orange ``Stellar'' curve in Figure~\ref{fig:feedback}, with column densities multiplied by a factor of $1.9$ to account for the differing UVBs.

The AGN feedback (included in the Illustris simulation shown with a green line)  suppresses the HI CDD at column densities $\NHI = 10^{12.5}-10^{14.5}$~\NHunit~by a factor of 1.5 with respect to the no feedback case.
This occurs not because of a change in the photoionization equilibrium\footnote{While the local radiation effects of the AGN are 
included, they are not large enough.}, but because the AGN both reduces the gas density at distances of $\sim 1$ Mpc  or more from galaxies and heats the remaining gas.
Both of these effects conspire to reduce the neutral gas density \citep{Sijacki:2015,Suresh:2015}. 
Figure \ref{fig:agn} shows the sensitivity of this effect to the strength of the AGN feedback. Here we alter the fraction of the AGN accretion 
energy, which couples to the gas via the radio mode (effective at suppressing star-formation).
Changing this tunable parameter by a factor of two with respect to the default setting used in Illustris changes the CDD by about $30 \%$, 
\cite{Genel:2014} showed that the Illustris AGN
feedback model is too strong as it over-suppresses the gas fraction in massive halos. However, the fact that varying the AGN feedback alters the
CDD suggests that AGN feedback models could now be tuned using the Lyman-$\alpha$ Forest. 
We therefore conclude that AGN feedback can have a significant effect on the CDD at low redshifts and discuss the implications of this further in Section \ref{sec:dis}.

\section{Discussion}
\label{sec:dis}

\subsection{Is there a ``Photon Underproduction Crisis?"}

\begin{figure*}
\includegraphics[width=0.46\textwidth]{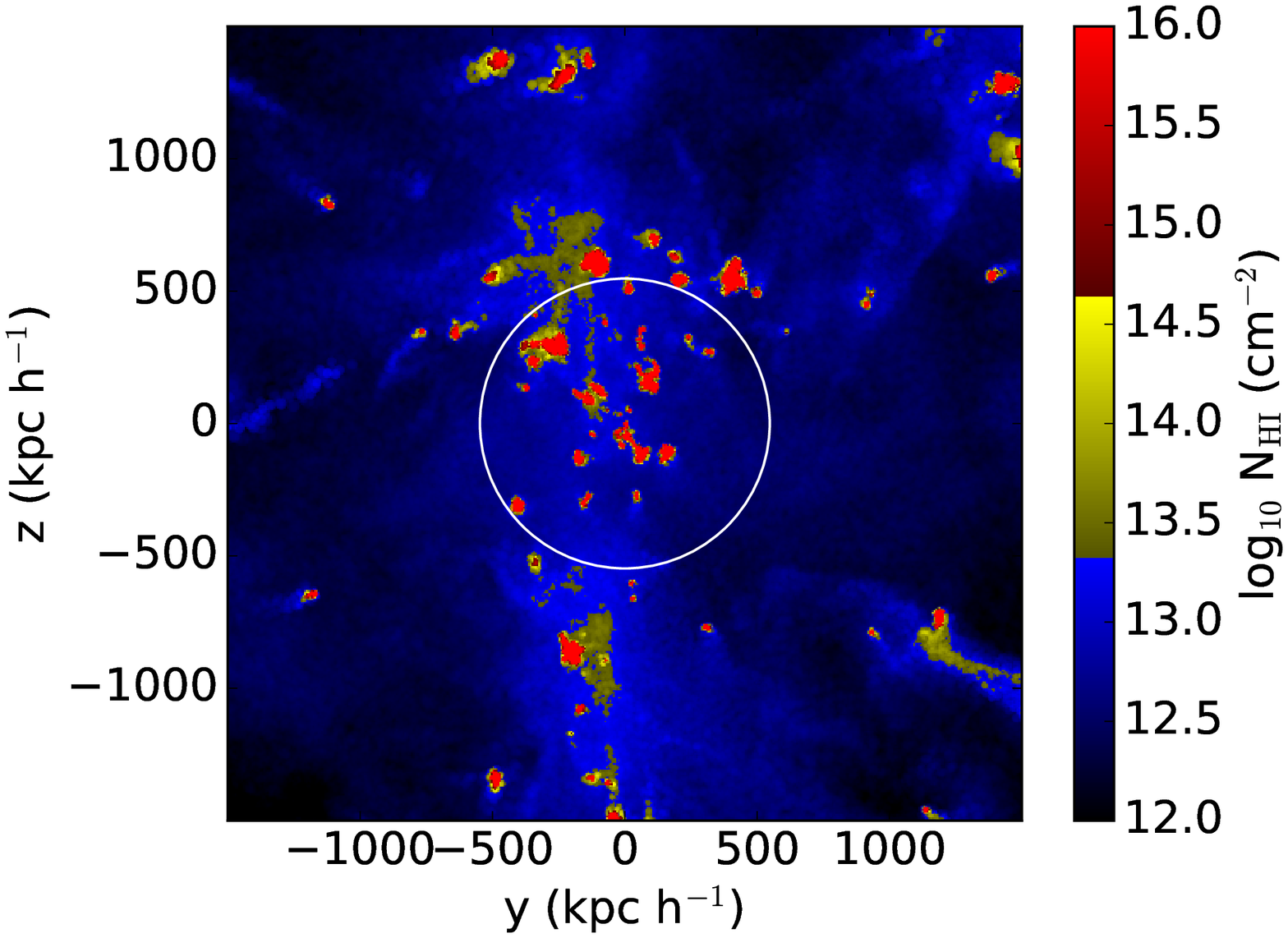}
\includegraphics[width=0.46\textwidth]{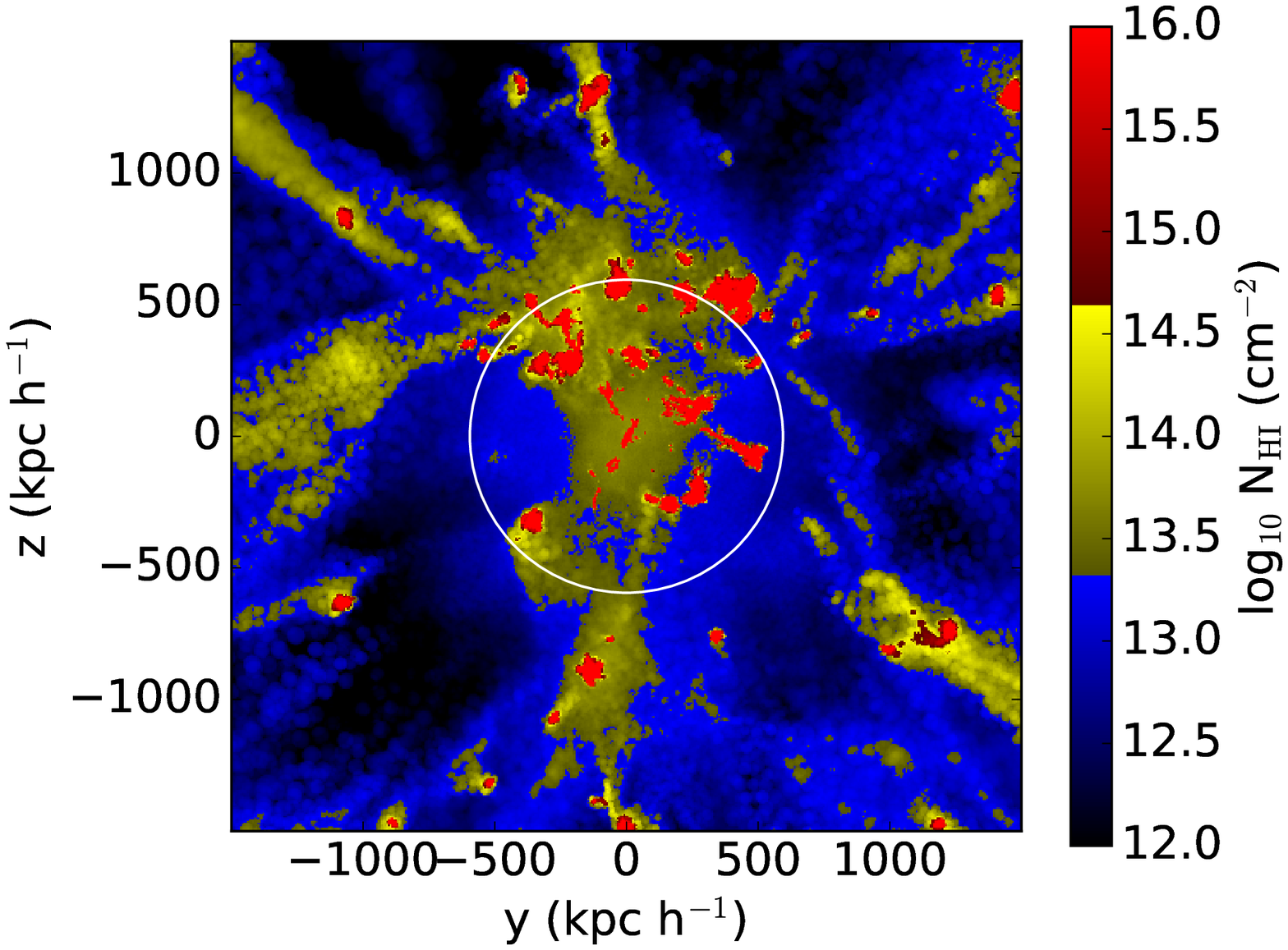}
\caption[ ]{
HI column density integrated over a $1.5$Mpc slice in a $1.5$Mpc box around the largest halo in our Illustris-small and Stellar simulations. 
(Left) Illustris-small simulation, which includes AGN and stellar feedback.
(Right) Stellar simulation, which does not include AGN feedback. The AGN feedback substantially reduces the HI column density 
up to $1.5$Mpc from the halo.
}
\label{fig:slice}
\end{figure*}

Recent studies have found that the UVB $\Gamma_{HI}$ model of \cite{HaardtMadau:2012} is not consistent 
 with the low-redshift CDD.  In particular, the predicted $z=0$ $\Gamma_{HI}$ from \cite{HaardtMadau:2012}~is five times lower than required by the simulations of K14
and three times lower than required by \cite{Shull:2015}. This has led to the suggestion of 
a  ``photon underproduction crisis'' for the low redshift \Lya~forest. K14 provided a detailed discussion of possible 
resolutions to the crisis, including boosting the quasar emissivity, the galaxy escape fraction, extra heating sources, altering the mean free path of ionizing
photons, as well as considering new sources of ionizing photons.

In this work, we investigate the effects of altering aspects of cosmological simulations including the feedback prescription, $\Gamma_{HI}$, and the hydrodynamic 
solver. We find that the main drivers, in our analysis, of the discrepancy between the D16 observational data and simulations are the absence of AGN feedback and the 
cosmic UV background model adopted for $\Gamma_{HI}$.  AGN feedback was not included in the GADGET simulations presented in K14.  We find that the inclusion of AGN 
feedback can substantially alter the CDD at the low column densities studied here.  This is because strong radio AGN feedback can substantially heat and ionize the gas, 
even more than $1$~Mpc from the halo \citep{Zhu:2016}, which will reduce the neutral column density. In the Illustris model, radio-mode AGN feedback leads to significant gas 
heating at $z<2$, especially in more massive halos, where it is the main regulator of star-formation \cite[see][]{Vogelsberger:2014}.
This is shown visually in Figure \ref{fig:slice}, where we plot the HI column density integrated in $1.5$ Mpc around the largest halo in our Illustris-small and Stellar 
simulations, which differ only in the presence of AGN feedback. AGN reduce HI densities, especially for the column ranges observed by \cite{Danforth:2016}. Stronger 
columns appear to be less affected, likely because higher density absorbers are harder to disrupt.
Our study is complimentary to those of \cite{Shull:2015}, \cite{Khaire:2015}, and \cite{Gaikwad:2016}, who showed that updated QSO and galaxy emissivity properties were sufficient to 
increase $\Gamma_{HI}$ by a factor of $\approx 2$ and could increase $\Gamma_{HI}$ by a factor of five with larger galaxy photon escape fractions.
While this suffices to produce enough photons, we show that with the addition of AGN feedback only a factor two increase in photon production is necessary, 
which is achieved, within current observational uncertainties, by using the FG09 model.
Therefore, with AGN feedback and the FG09 UVB, Illustris is able to better match the D16 observations for $\NHI = 10^{12.5} - 10^{13.5}$~\NHunit and the ``photon underproduction crisis'' can be fully resolved.

\subsection{The AGN Feedback Model in Illustris and Higher Column Densities}

As shown in \cite{Genel:2014}, the Illustris AGN feedback model over-suppresses the gas fraction in massive halos, indicating that
the feedback model is overly effective at expelling gas from the halo. However, this problem is unlikely to affect the results we present 
for $\NHI = 10^{12.5} - 10^{14}$~\NHunit. These absorbers correspond to baryon over-densities in the range $4-40$ \citep{Shull:2015} and are thus around $2$Mpc from the halo.
For comparison, the halos in which the gas fraction is over-suppressed have masses $10^{13}-10^{14} M_\odot$ and virial radii $R_{500} = 300 - 700$ kpc.
They also comprise a small fraction of the total volume of the simulation, which is dominated by smaller halos.
Furthermore, a suppression of neutral gas at megaparsec distances from the halo is a generic feature of AGN feedback models \citep{vanDaalen:2011}. 
Overall, then, our principle result that AGN feedback is a significant factor in reconciling simulations with the D16 observations at 
low column densities seems robust to modest changes in the underlying model. 

There is a discrepancy between the COS observations and our AREPO simulations at $10^{13.5} < \NHI < 10^{15.5}$~\NHunit.
In this column density range, Illustris produces a factor of two too few absorbers.  These stronger columns are associated 
with denser regions closer to the halos, where the over-vigorous expulsion of gas in the Illustris AGN feedback model may 
have a larger impact. There may thus be sufficient freedom in the model to resolve the disagreement between observations and simulations.
This would not be the only example of a discrepancy between observations and Illustris resulting from the AGN feedback model.
For example, in the Illustris simulation the stellar masses of the central galaxies in the simulated systems are also too high 
as a result of the overactive AGN feedback model. Future studies will investigate the low-redshift \Lya~forest CDD using 
an updated model for AGN feedback \citep{Weinberger:2016} in order to investigate the number of absorbers at medium column densities.

We note, however, that our models with no feedback and supernova feedback also underproduce absorbers in 
the column density range $10^{14} < \NHI < 10^{15.5}$~\NHunit. K14 simulations do not suffer from this discrepancy, but their simulations did not include the effect of self-shielding, which 
may start to become important for these column densities. Interestingly, the simulations of \cite{Shull:2015} also underproduce 
stronger systems in their preferred model. This suggests that the full solution may be more involved.

For example, in the flat ($\NHI=10^{14}-10^{18}$~\NHunit) regime of the curve of growth (CoG), the HI 
equivalent width depends strongly on both the column density of the absorber and the kinematic and 
thermal state of the gas (i.e. the Doppler broadening of the line). Thus, the CDD could potentially be affected by turbulent broadening, which is unresolved
in numerical simulations. Indeed, most turbulence box simulations require grid resolutions of at least $512^3$ before resolving the inertial range cascade \citep{Burkhart:2009}.
Cosmological numerical simulations do not have the spatial resolution to resolve turbulence in the IGM and this is especially true of simulations 
that set the spatial refinement based on quasi-Lagrangian refinement, which puts most of the spatial resolution inside galaxies. Therefore, the IGM in Illustris, which has the poorest spatial resolution sampling because it 
consists of the lowest density environments, cannot resolve all of the kinematic motions that may be present 
in the observations. The effects of turbulence on the IGM statistics, including the \Lya~CDD, in the linear regime of the CoG will be closely examined in a future work.

We found that the CDD in this column density range was also affected by the hydrodynamic solver used.
SPH simulations can suppress turbulence and mixing, which can affect thermal conduction \citep{Biffi:2015} galaxy 
formation \citep{Keres:2012} in idealized test cases \citep{Sijacki:2012}. While the differences between \arepo~and 
\gadget~were small at lower column densities, they were a factor of two for $10^{14} < \NHI < 10^{15.5}$~\NHunit.
This mirrors the results of \cite{Bird:2013}, who found that SPH tends to produce spurious clumps in gas at this physical density.
Any future simulation work on this discrepancy should use a hydrodynamical solver that is sufficiently 
accurate in this regime.

\section{Conclusion}
\label{sec:con}

We examined the $z=0.1$ CDD in the Illustris simulation. We find that the Illustris simulation matches 
the D16 observations for the column density range of $\NHI = 10^{12.5} - 10^{13.5}$~\NHunit significantly better than previous work 
for the same photoionization rate as a result of the inclusion of AGN
feedback. We investigated the effects on our results of 
the hydrodynamic solver, supernova, and AGN feedback models, as well as the UVB model.
We found that the most significant factors affecting agreement with observations were the UVB model and the inclusion of 
AGN feedback (radio mode, quasar mode, and radiative). The \cite{Faucher:2009} UVB model we used, in conjunction with the inclusion of AGN feedback, 
matches current observations at low redshifts at a better level than \cite{Kollmeier:2014} simulations that used 
the UVB of \cite{HaardtMadau:2012} and no AGN feedback. However, in contrast to the situation at $z>2$, our simulations underproduce absorbers 
at column densities $\NHI > 10^{13.5}$~\NHunit, which may indicate a need to modify feedback models.

We showed that AGN feedback can significantly suppress the column density function in the range of $\NHI = 10^{12.5} - 10^{14.5}$~\NHunit.
The effect of AGN feedback has not formerly been addressed and so our work demonstrates a new potential solution to the ``photon underproduction crisis.''
While the AGN feedback model in Illustris needs to be refined,  our results suggest that AGN feedback can significantly affect these column densities, and 
thus must be considered in any future work on the HI CDD. The measurements of \cite{Danforth:2016} may constitute the first 
observational evidence for the effect of AGN feedback on neutral gas around galactic halos and could provide an additional
diagnostic to tune AGN feedback models in simulations.

\appendix

\section{Numerical Convergence}

\begin{figure}
  \centering
  \begin{minipage}[b]{0.45\textwidth}
  \includegraphics[width=\textwidth]{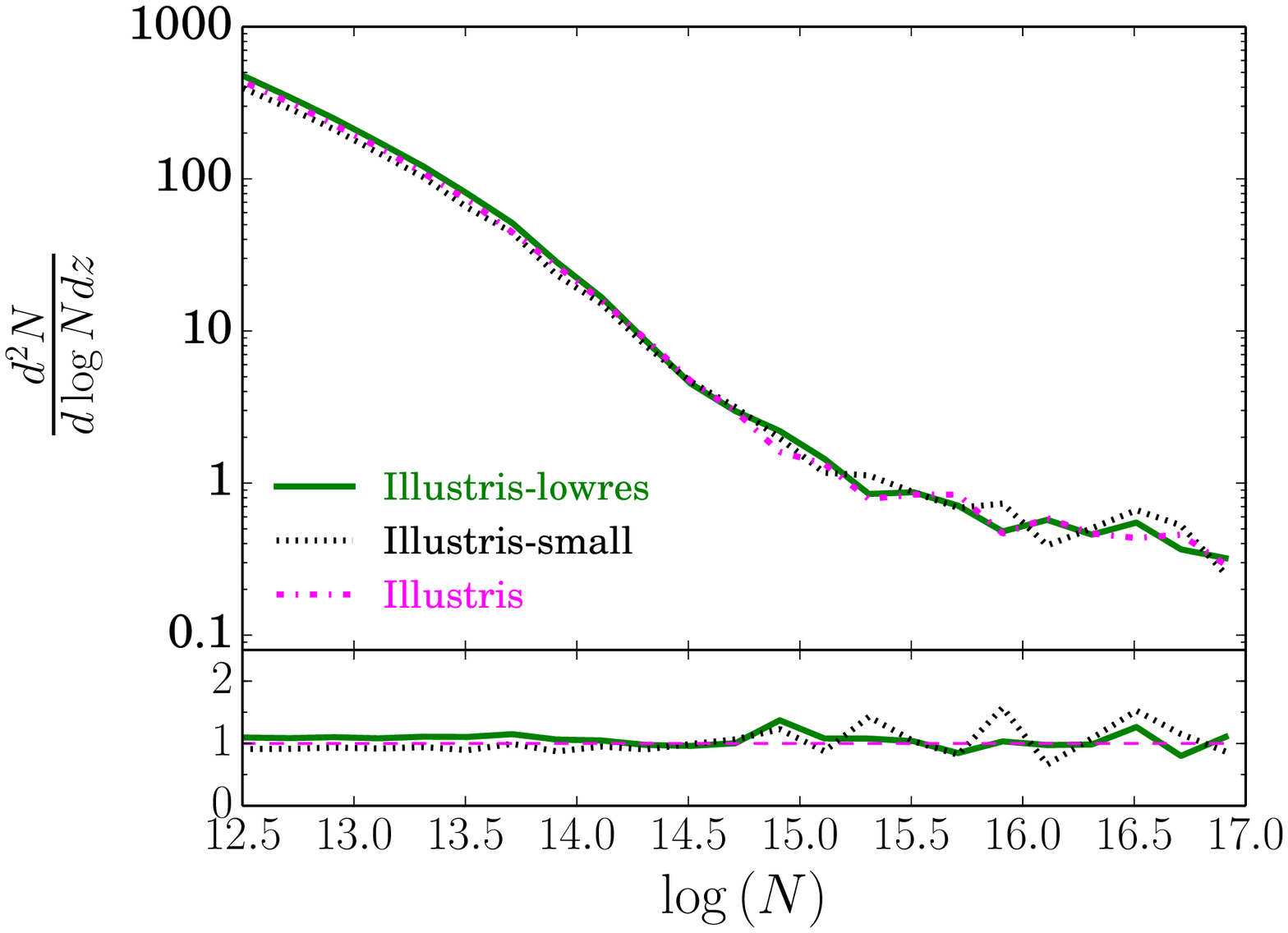}
    \caption[ ]{HI CDD as measured using the direct method at $z=0.1$ from the full Illustris simulation (magenta dot-dashed), 
    with a $75$ Mpc box and $1820^3$ particles, compared to Illustris-small (black dotted), a $25$ Mpc box with $512^3$ particles. 
    Also shown is Illustris-lowres (green solid), a $25$ Mpc box with $256^3$ particles.
The lower panel shows the ratio of each simulation to Illustris.
}\label{fig:resolution}
  \end{minipage}
  \hfill
  \begin{minipage}[b]{0.45\textwidth}	
    \includegraphics[width=\textwidth]{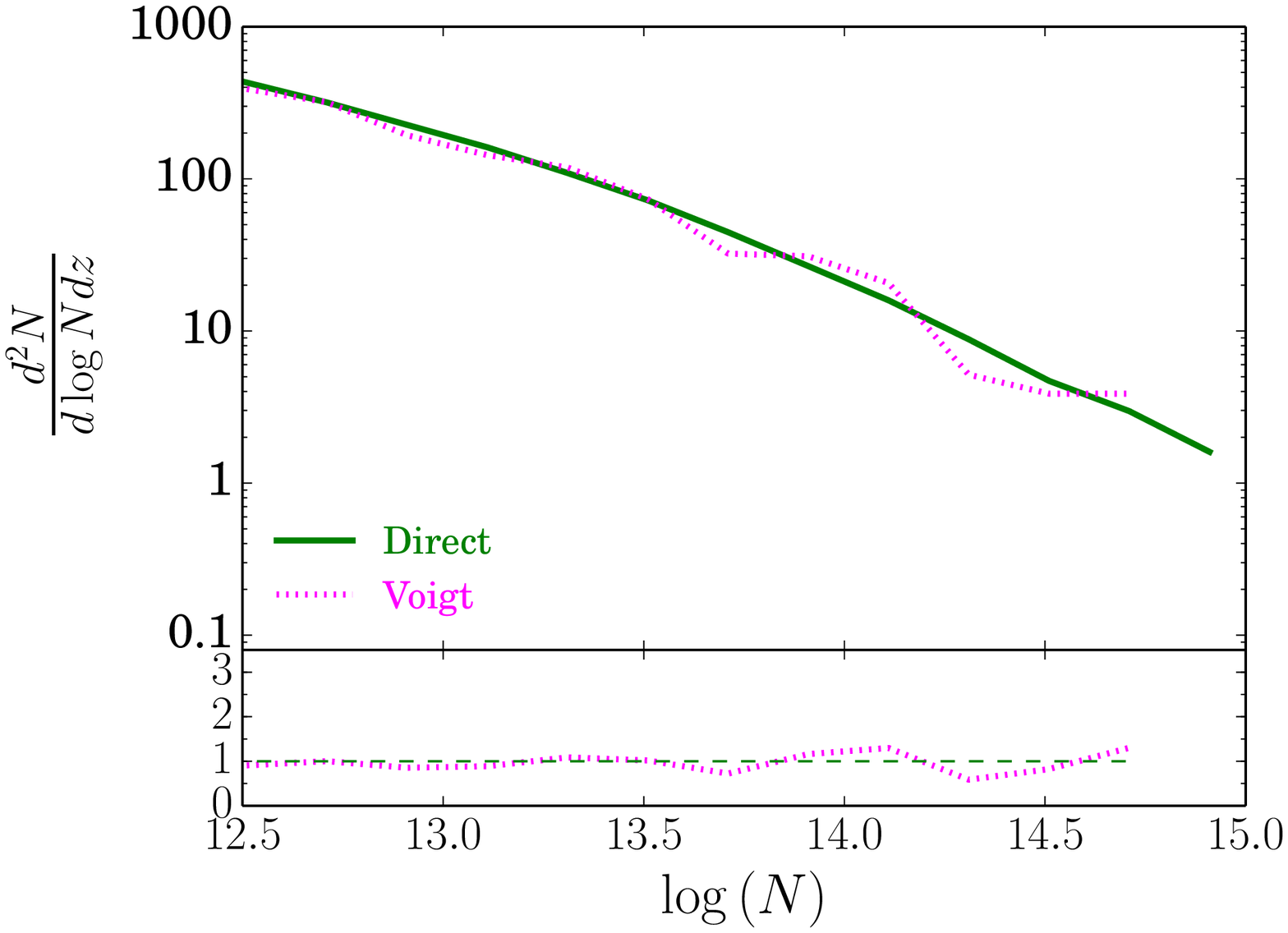}
\caption[]{Column density function from Illustris $75$ Mpc at $z=0.1$, estimated using both
direct summation of column densities (green solid) and using Voigt profile fitting (magenta dotted).
The lower panel shows the ratio of the Voigt fitted CDD to the direct summation estimate.
}
  \end{minipage}
  \label{fig:voigt}
\end{figure}

%
%
%
%

Figure~\ref{fig:resolution} shows a convergence test of our simulations, examining both resolution and box size.
We compare the full $75 \Mpch$ Illustris box with $1820^3$ particles to identical simulations in a $25 \Mpch$ box. The first 
simulation (Illustris-small) has $2\times 512^3$ resolution elements, and thus identical spatial resolution to Illustris, in a smaller box. 
The second simulation (Illustris-lowres) has $2\times 256^3$ resolution elements, and thus a factor of two lower spatial resolution. 
Differences between the CDDs in the column density range of interest are negligible and due mostly to sample variance, demonstrating
that our results are insensitive to changes in both the box size and the resolution.

\section{Voigt Fitting}
\label{sec:voigt}

\begin{figure}
\centering
    \includegraphics[width=0.4\textwidth]{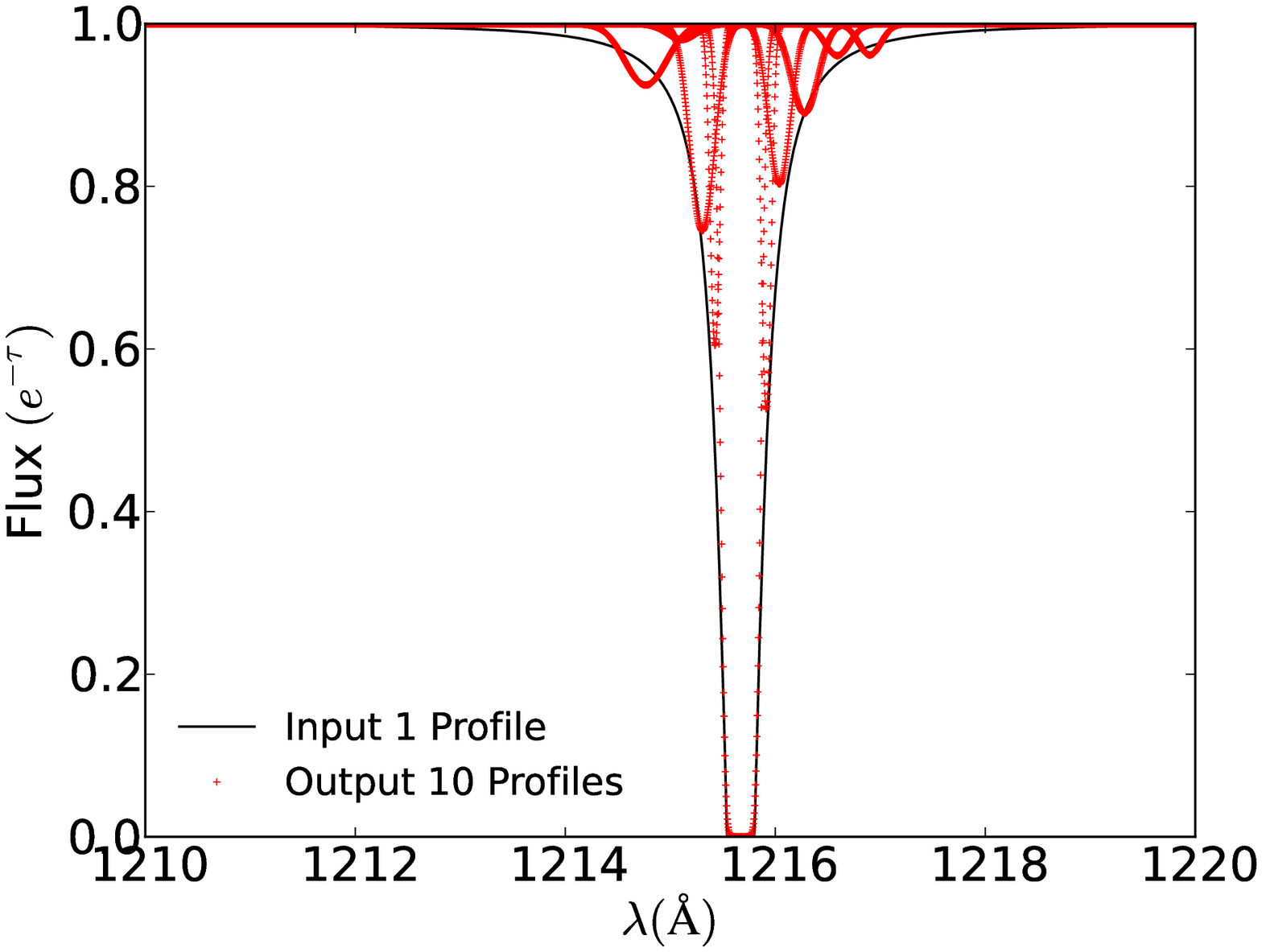}
  \includegraphics[width=0.4\textwidth]{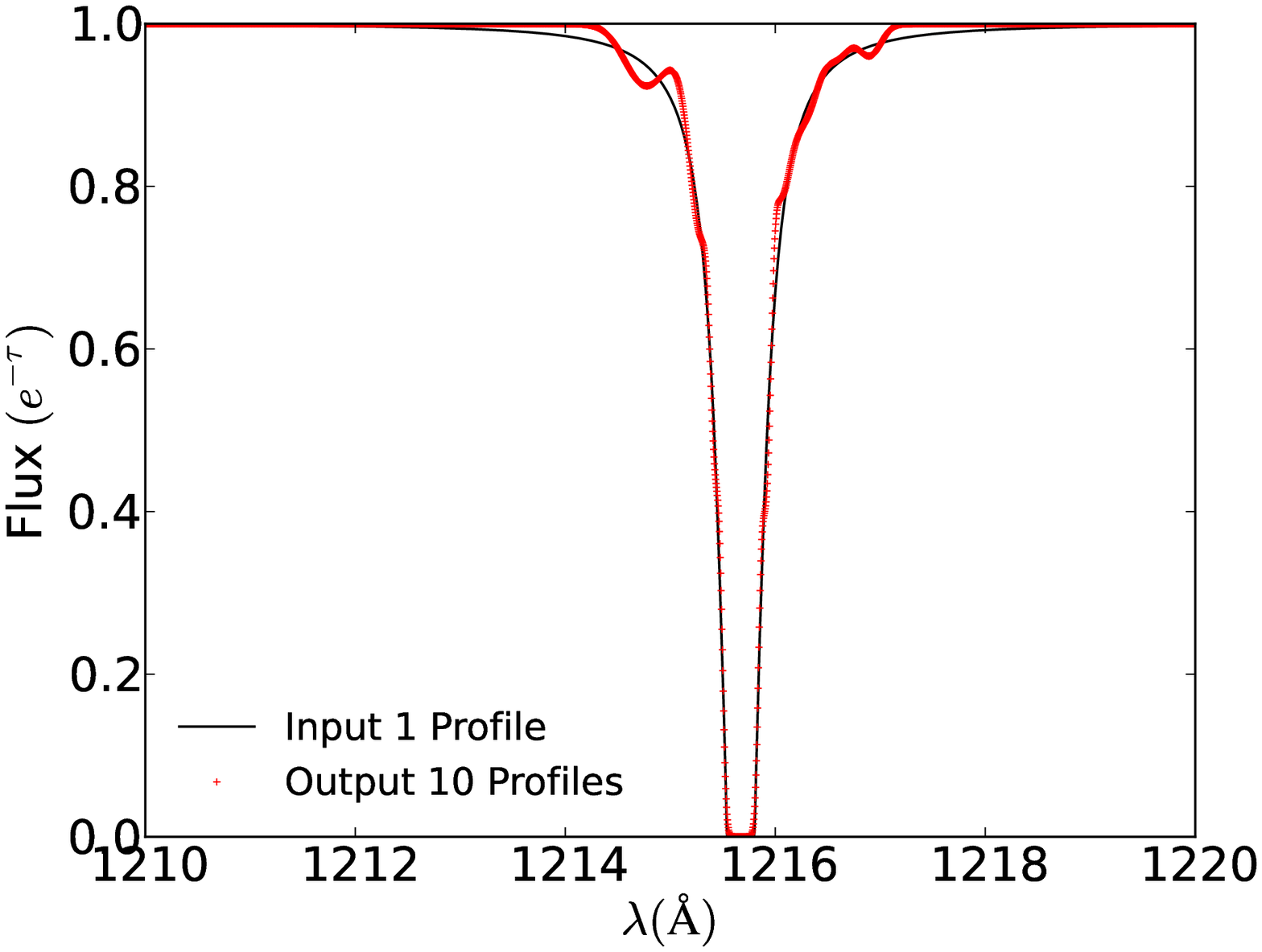}
  \caption{\label{fig:autovp} (Left) Idealized Voigt profile (black) with $b=10 \kms$ and $N=10^{18}$ \NHunit with the output parameters of the fit Voigt profiles 
from \url{AUTOVP} stacked on top (red). (Right) Same Voigt profile (black) overlaid with the sum of the fitted output profiles (red).
}
 \end{figure}

We have estimated the CDD using two separate methods. The first, which is used for all 
figures in the main section of the paper, we call direct summation. The column density here is estimated via the integral of the 
density field in $50$~\kms along a sightline. The bins exclude peculiar velocities and reflect the Hubble flow. The CDD function is computed via a histogram 
of the column densities. This column density estimator has the advantage of being extremely quick and scalable. However, it differs from 
the practice of observers, who use Voigt profile fitting to estimate the column density from optical depth spectra.

In this appendix, we will estimate the CDD using automated Voigt fitting as a check that this difference does not affect our results.
First, artificial spectra are generated from each particle. Each particle is redshifted according to its peculiar velocity and the optical depth is generated 
by convolving a Voigt profile with the SPH kernel. Thermal broadening is included using the temperature of the particle.
We then estimate the column density that would be observed using an automated Voigt profile fitter on these simulated spectra.
Following \cite{Kollmeier:2014}, we used primarily \begin{small}AUTOVP\end{small}\footnote{\url{https://bitbucket.org/benopp/autovp_phys}} \citep{Dave:1997}.
AUTOVP is a fully automated Voigt fitter that fits the profiles after adding simulated noise. Each line is subtracted in turn until the 
remaining features are consistent with the noise in the spectrum. A global re-fit of the lines is then performed. 
We used the default \small{AUTOVP} parameters and verified that our results were independent of adjusting these values.

Figure~\ref{fig:voigt} shows the CDD estimated from both Voigt fitting and direct summation. They are in extremely good agreement. 
The resulting minor discrepancies are consistent with variance due to the added noise in the Voigt spectra. Note that due 
to the increased computational cost of Voigt fitting, Figure~\ref{fig:voigt} is estimated using only $5000$ sightlines.
This validates our decision to use direct summation for the main results in our paper.

Figure~\ref{fig:voigt} does not include results for column densities $\NHI > 10^{15}$~\NHunit. In this partially saturated regime we were unable to 
reliably fit the profiles with the publicly available version of \small{AUTOVP}. Figure~\ref{fig:autovp} shows an example of this problem on an 
idealized test case. While the overall fit to the profile is visually reasonable, \small{AUTOVP} is arbitrarily fitting the high column density 
absorber with multiple lower column density systems. Fortunately, high column density systems are rare, and thus this problem does not 
substantially impact lower column density absorbers. We have verified this point using our own parallel Voigt fitter implemented in 
Python\footnote{\url{https://github.com/sbird/fake_spectra/voigtfit.py}}, which we verified is able to correctly fit high column density systems.
The resulting CDD still matches that estimated by direct summation, including at column densities up to $10^{17}$~\NHunit.

\acknowledgments
The authors are grateful for many valuable discussions with Lars Hernquist.
The authors also thank the anonymous reviewer and Claude-Andr\'e Faucher-Gigu\`ere for their very helpful suggestions.  
A.G. acknowledges support from the NSF REU program through grant Number 1262851. 
B.B. acknowledges support from the NASA Einstein Postdoctoral Fellowship. 
S.B. was supported by NASA through Einstein Postdoctoral Fellowship Award Number PF5-160133.

\bibliography{ms}

\end{document}